\documentclass[letterpaper,prl,twocolumn,showpacs]{revtex4}%
\usepackage{graphicx}
\usepackage{times}
\begin{document}

\title{An experimental investigation of criteria for continuous
variable entanglement}

\author{W.~P.~Bowen, R.~Schnabel, and P.~K.~Lam}

\address{Department of Physics, Faculty of Science, Australian 
National University, ACT 0200, Australia}

\author{T.~C.~Ralph}

\address{Department of Physics, Centre for Quantum Computer
Technology, University of Queensland, St Lucia, QLD 4072, Australia}

\begin{abstract}
We generate a pair of entangled beams from the interference of two
amplitude squeezed beams.  The entanglement is quantified in terms of
EPR-paradox \cite{Reid88} and inseparability\cite{Duan00} criteria,
with observed results of $\Delta^{2}\!  X_{x|y}^{+} \Delta^{2}\! 
X_{x|y}^{-} \!  = \!  0.58 \!  \pm \!  0.02$ and $\sqrt{\Delta^{2}\! 
X_{x \pm y}^{+} \Delta^{2}\!  X_{x \pm y}^{-}}\!  = \!  0.44 \!  \pm
\!  0.01$, respectively.  Both results clearly beat the standard
quantum limit of unity.  We experimentally analyze the effect of
decoherence on each criterion and demonstrate qualitative differences. 
We also characterize the number of required and excess photons
present in the entangled beams and provide contour plots of the
efficacy of quantum information protocols in terms of these variables.
\end{abstract}
\pacs{42.50.Dv, 42.65.Yj, 03.67.Hk}
\maketitle
{\it Entanglement} is one of the most interesting properties of
quantum mechanics, and is an important ingredient of quantum
information protocols such as quantum teleportation\cite{Bennett93},
densecoding \cite{Bennett92, Ralph02} and quantum computation
\cite{DiVincenzo95}.  In the Schr\"odinger picture, a necessary and
sufficient criterion for the emergence of entanglement is that the
state describing the entire system is {\it inseparable}, i.e. the
wavefunction of the total system cannot be factored into a product of
separate contributions from each sub-system.  Using the Heisenberg
approach, a sufficient criterion for the presence of entanglement is
that correlations between conjugate observables of two sub-systems
allow the statistical inference of either observable in one
sub-system, upon a measurement in the other, to be smaller than the
standard quantum limit, i.e. the presence of non-classical
correlations.  The latter approach was originally proposed in the
paper of Einstein, Podolsky and Rosen \cite{Einstein35}.  These two
different pictures result in two distinct methods of characterizing
entanglement.  One is to identify an observable signature of the
mathematical criterion for wave-function entanglement, i.e.
inseparability of the state.  The second looks directly for the onset
of non-classical correlations.  For pure states these two approaches
return the same result suggesting consistency of the two methods. 
However, when decoherence is present, causing the state to be mixed,
differences can occur.

We present the generation and investigation of Gaussian continuous
variable entanglement between the amplitude and phase quadratures of a
pair of light beams; henceforth termed {\it quadrature entanglement}. 
We will use the measure proposed by Duan {\it et al.} \cite{Duan00} as
an example that identifies wave-function inseparability and refer to
this as the {\it inseparability criterion}.  As an example of an
entanglement criterion based directly on the observation of
non-classical correlations, we will use the demonstration of the
EPR-paradox as quantified by Reid and Drummond \cite{Reid88}, and
refer to this as the {\it EPR criterion}.  We introduce decoherence in
the form of optical loss to both entangled beams and experimentally
demonstrate qualitative and quantitative differences between the two
criteria.  These differences suggest a more comprehensive
representation may be appropriate.  We represent the entanglement on a
plot of the number of photons required to generate it versus the
number of excess photons.  We present efficacy contours for some
common quantum information protocols on this plot.

\begin{figure}[t]
    \begin{center}
    \includegraphics[width=7.5cm]{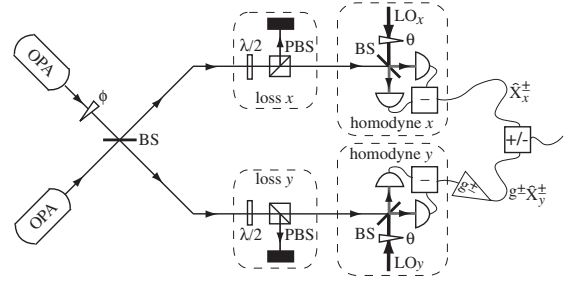}
    \end{center}
    \vspace{-5mm} \caption{Experimental schematic.  BS (PBS): 50/50
    (polarizing) beam splitter,
    $\lambda/2$: half-wave plate, $\phi$ and $\theta$: phase shift.}
    \label{Expt}
\end{figure}
Fig.~\ref{Expt} shows the experimental setup that was used to generate
quadrature entanglement.  Two independent amplitude squeezed beams at
1064~nm were produced by a pair of type I optical parametric
amplifiers (OPAs) constructed from hemilithic MgO:LiNbO$_{3}$ crystals
and output couplers.  The seed power to the OPAs was adjusted so that
the two squeezed beams were of equal intensity.  Both had 4.1~dB of
squeezing at 6.5~MHZ as measured in a homodyne detector with total
detection efficiency of 85~\%.  Further details of this setup are
reported in \cite{Bowen02}.  We produced a pair of quadrature
entangled beams by interfering the squeezed beams with relative phase
of $\pi/2$ on a 50/50 beam splitter, and observed a visibility of
$98.7 \!  \pm \!  0.3$~\% for the process.  Each of the entangled
beams was interrogated in a balanced homodyne detector that could be
locked to detect either its phase or amplitude quadrature.  We
observed $96 \!  \pm \!  0.5$~\% visibility between each entangled
beam and its homodyne local oscillator.  Epitaxx ETX500 photodiodes
with approximately 93~\% quantum efficiency were used.  Both the EPR
and inseparability criteria could be quantified by analyzing
correlations between these two homodynes.

The EPR criterion for a given pair of beams $x$ and $y$ can be related
to their degree of quantum correlation.  This is quantified by the
product of conditional variances $\Delta \!  ^{2} \!  \hat X \!_{x|y}$
of conjugate quadratures between the beams.  Limiting ourselves to the
amplitude $\hat X^{+}$ and phase $\hat X^{-}$ quadratures, the EPR
criterion is \cite{Reid88}
\begin{equation}
	\label{VcvVcv}
\Delta \!  ^{2} \!  \hat X \!^{+}_{x|y} \Delta \!  ^{2} \!  \hat X
\!^{-}_{x|y} < 1
\end{equation}
where $\Delta \!  ^{2} \!  \hat X \!^{\pm}_{x|y} \!  = \!  \Delta \! 
^{2} \!  \hat X \!^{+}_{x} \!  - \!  | \langle \delta \!{X}^{\pm}_{x}
\delta \!  {X}^{\pm}_{y} \rangle |^{2}/ \Delta \!  ^{2} \!  \hat X
\!^{\pm}_{y} \!  = \!  \min_{g^{\pm}}{\langle ( \delta \!{X}^{\pm}_{x}
\!  - \!  g^{\pm} \delta \!{X}^{\pm}_{y})^{2} \rangle}$; we expand the
quadrature operators into a mean unchanging term $\langle \hat
X^{\pm} \rangle$ and a noise operator $\delta \hat X^{\pm}$, $\hat
X^{\pm} \!  = \!  \langle \hat X^{\pm} \rangle \!  + \!  \delta \hat
X^{\pm}$.  Satisfaction of the EPR criterion between two beams is a
sufficient but not necessary condition for their entanglement (i.e
inseparability).  It is easily measurable, and has been used to
characterize entanglement in a number of experiments
\cite{conditional}.

We are interested in the effect of decoherence in the form of optical
loss on the EPR and inseparability criteria.  For entanglement
generated as in fig.~\ref{Expt} from a pair of pure squeezed beams
both with squeezed variances $\Delta \!  ^{2} \!  \hat X \!_{{\rm
sqz}}$, the left-hand-side of eq.~(\ref{VcvVcv}) as a function of
detection efficiency $\eta$ is
\begin{eqnarray}
\label{VcvVcvloss}
\Delta \!  ^{2} \!  \hat X \!^{+}_{x|y} \Delta \!  ^{2} \!  \hat X
\!^{-}_{x|y} \!  = \!  4 \!  \left ( \!  1 \!  - \!  \eta \!  + \! 
\frac{2 \eta \!  - \!  1}{\eta (\Delta \!  ^{2} \!  \hat X \!_{{\rm
sqz}} \!  + \!  \Delta \!  ^{2} \! \hat X \!_{{\rm sqz}}^{-1} \!  -
\!  2) \!  + \!  2} \!  \right )^{\hspace{-0.8mm} 2}
\end{eqnarray}
Notice that when $\eta \!  = \!  0.5$, $\Delta \!  ^{2} \!  \hat X
\!^{+}_{x|y} \Delta \!  ^{2} \!  \hat X \!^{-}_{x|y} \!  = \!  1$,
independent of the level of squeezing; it follows that for $\eta \!  <
\!  0.5$ the EPR criterion cannot hold.

The inseparability criterion relies on the identification of
separability with positivity of the P-representation distribution of
the state.  For states with Gaussian noise distributions and symmetric
correlations on the conjugate quadratures, it can be related to
measurable correlations \cite{Duan00}
\begin{equation}
\bigg \langle \!  \Big (|a| \hat X_{x}^{+} \!  +   {{\hat
X_{y}^{+}}\over{a}} \Big )^{\hspace{-0.8mm} 2} \bigg \rangle \! + \! 
\bigg \langle \!  \Big (|a| \hat X_{x}^{-} \!  -   {{\hat
X_{y}^{-}}\over{a}} \Big )^{\hspace{-0.8mm} 2} \bigg \rangle < 2 \! 
\left ( \!  a^{2} \!  + \!  {{1}\over{a^{2}}} \!  \right )
\label{Duangeneral}
\end{equation}
where $a$ is an experimentally adjustable parameter.  In our
experiment the entangled beams were produced on a 50/50 beam splitter
and were indistinguishable from each other; in this case $a \!  = \! 
1$ and the criterion can be written $ \Delta \!  ^{2} \!  \hat X
\!^{+}_{x \pm y} \!  + \!  \Delta \!  ^{2} \!  \hat X \!  ^{-}_{x \pm
y} \!  < \!  2$, where $\Delta \!  ^{2} \!  \hat X_{x \pm y}$ is the
minimum of the variance of the sum or difference of the operator $\hat
X$ between beams $x$ and $y$ normalized to the two beam shotnoise,
$\Delta \!  ^{2} \!  \hat X_{x \pm y} \!  = \!  \min{\langle ( \delta
\hat X_{x} \!  \pm \!  \delta \hat X_{y})^{2} \rangle}/2$.  The
symmetric correlation restriction on eq.~(\ref{Duangeneral}) becomes
unnecessary for indistinguishable entangled beams when the criterion
is written in a product form
\begin{equation}
\label{Duan}
\sqrt{\Delta \!  ^{2} \!  \hat X \!^{+}_{x \pm y}  \Delta \!  ^{2} \!
\hat X \!  ^{-}_{x \pm y}} < 1
\end{equation}
For our experimental configuration the left-hand-side of
eq.~(\ref{Duan}) as a function of efficiency can be expressed
\begin{equation}
    \label{Duanloss}
\sqrt{\Delta \!  ^{2} \!  \hat X \!^{+}_{x \pm y} \Delta \!  ^{2} \! 
\hat X \!  ^{-}_{x \pm y}} = \eta \Delta \!  ^{2} \!  \hat X \!_{{\rm
sqz}} + (1 - \eta)
\end{equation}
Notice that, in contrast to the EPR criterion, even as $\eta$
approaches zero, for any level of squeezing eq.~(\ref{Duanloss}) is
below unity.  It can be shown from
eqs.~(\ref{VcvVcvloss})~and~(\ref{Duanloss}) however, that for $\eta
\!  = \!  1$, (i.e. for pure measured states)
criteria~(\ref{VcvVcv})~and~(\ref{Duan}) are qualitatively equivalent.

\begin{figure}[b]
    \begin{center}
    \includegraphics[width=8.7cm]{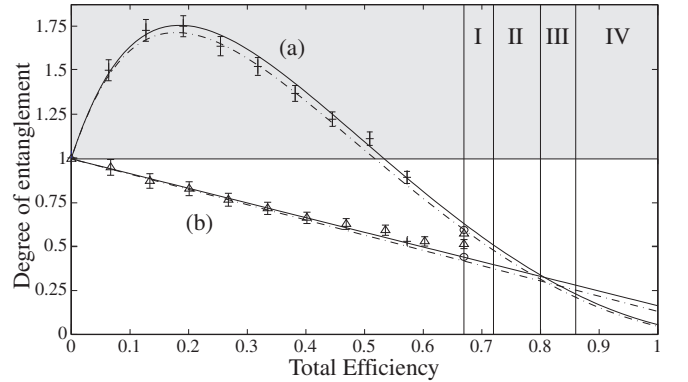}
    \end{center}
    \vspace{-5mm} \caption{Comparison of (a) EPR and (b)
    inseparability criteria with varied detection efficiency.  The
    symbols $+$, $\triangle$, and {\small $\bigcirc$} label three
    separate experimental runs.  For $+$ a systematic error was
    introduced by the detection darknoise when optimizing the EPR
    criterion gain.  The solid fit in (a) includes this, the dashed
    fit is the result expected if the error was eliminated, and agrees
    well with runs $\triangle$ and {\small $\bigcirc$}.  The solid
    line in (b) is a theoretical fit, the dashed line is the result
    predicted by the fit in (a).  There were four sources of
    unavoidable loss in our system, I: Detection
    loss, II: Homodyne loss, III: optical loss and IV: OPA escape
    loss.}
    \label{results}
\end{figure}
In our experiment correlations were observed between the output
currents from the homodyne detectors measuring beams $x$ and $y$ when
locked simultaneously to the amplitude or phase quadrature.  The
inseparability criterion was determined by analyzing the unity gain
sum and difference photocurrents between the two homodynes in a
spectrum analyzer.  The spectrum analyzer had 300~kHz resolution
bandwidth and 300~Hz video bandwidth and was set to zero span at
6.5~MHz.  Ten consecutive traces each containing 400 measurement
points were averaged for each result.  All measured traces were at
least 4.5~dB above the detection darknoise which was taken into
account.  The variances of the sum and difference photocurrents were
normalized to the vacuum noise scaled by the combined power of the two
homodyne local oscillators and the two entangled beams.  The minimum
of these two normalized variances is $\Delta \!  _{x \pm y}^{2} \! 
\hat X$ for the quadrature being measured in the homodynes.  With no
additional loss in the system we observed $\Delta \!  ^{2} \!  \hat X
\!^{\pm}_{x \pm y} \!  = 0.44 \!  \pm \!  0.01$ for both the amplitude
and phase quadratures.  The left-hand-side of criterion (\ref{Duan})
was then $\sqrt{\Delta \!  ^{2} \!  \hat X \!^{+}_{x \pm y} \Delta \! 
^{2} \!  \hat X \!^{-}_{x \pm y}} \!  = \!  0.44 \!  \pm \!  0.01$
which is well below unity and indicates that our beams were indeed
quadrature entangled.

The EPR criterion was determined in a similar manner to that described
above.  This time, however, rather than being fixed to unity, the gain
between the two homodyne photocurrents was optimized to minimize the
measured variances; and the normalization was performed with respect
to vacuum fluctuations scaled by only one homodyne local oscillator
and entangled beam.  The optimum results obtained for each quadrature
were $\Delta \!  ^{2} \!  \hat X \!^{+}_{x|y} \!  = \!  0.77 \!  \pm
\!  0.01$ and $\Delta \!  ^{2} \!  \hat X \!^{-}_{x|y} \!  = 0.76 \! 
\pm \!  0.01$, and the left-hand-side of criterion~(\ref{VcvVcv}) was
in the regime for observation of the EPR-paradox, $\Delta \!  ^{2} \! 
\hat X \!^{+}_{x|y} \Delta \!  ^{2} \!  \hat X \!^{-}_{x|y} \!  = \! 
0.58 \!  \pm \!  0.02 \!  < \!  1$.

As discussed earlier, the criterion for demonstration of the
EPR-paradox and inseparability are qualitatively different.  In
particular, the criterion for demonstration of the EPR-paradox
strongly depends on mixedness, whereas the inseparability criterion is
independent of it.  We demonstrated this experimentally by adding loss
to each of the quadrature entangled beams.  This changed both the
degree of entanglement and also it's mixedness.  Placing a polarizing
beam splitter and half-wave plate in each entangled beam allowed the
introduction of arbitrary loss.  We took measurements of both criteria
for a number of loss settings, the results are shown in
fig.~\ref{results}.  The experiment agrees very well with our
theoretical predictions.  As discussed in \cite{Duan00}, no matter
what the loss, the inseparability criterion always holds.  The EPR
criterion however, fails for loss greater than 0.48.  In fact as
observed earlier, it is impossible for the EPR criterion to hold for
loss greater than or equal to 0.5.  The error bars on the plots can be
attributed to uncertainty in the loss introduced, small fluctuations
in the local oscillator powers and, for the EPR criterion, error in
the optimization of the electronic gain.

Many applications have been proposed for entanglement.  In general,
these applications differ in their sensitivity to the mixedness of the
entanglement.  For example, unity gain quantum teleportation is
independent of mixedness, whereas densecoding can be very sensitive to
it.  For this reason, it is useful to characterize entanglement on a
two dimensional diagram which includes both the degree of entanglement
and its mixedness.  This may be conveniently achieved with a photon
number diagram.

The average number of photons per unit bandwidth per unit time in a
sideband of an optical beam is given by $\bar n \!  = \!  \langle
\delta \hat a^{\dagger} \delta \hat a \!  \rangle$, where $\hat a$ is
the field annihilation operator and can be related to the quadrature
operators by $\hat a \!  = \!  (\hat X^{+} \!  + \!  i \hat X^{-})/2$. 
$\bar n$ can be written in terms of quadrature variances as $\bar n \! 
= \!  (\Delta \!  ^{2} \!  \hat X^{+} \!  + \!  \Delta \!  ^{2} \! 
\hat X^{-} \!  - \!  2)/4$.  In general some minimum number of photons
$\bar n_{{\rm min}}$ is required to maintain a given degree of
entanglement between a pair of optical beams.  For indistinguishable
entanglement with Gaussian noise statistics $\bar n_{{\rm min}}$ can
be expressed as a function of the left-hand-side of the inseparability
criterion in eq.~(\ref{Duan})
\begin{equation}
\bar n_{{\rm min}} \!  = \!  \frac{(\Delta \!  ^{2} \!  \hat X
\!^{+}_{x \pm y} \Delta \!  ^{2} \!  \hat X \!  ^{-}_{x \pm
y})^{\frac{1}{2}} \!  + \!  (\Delta \!  ^{2} \!  \hat X \!^{+}_{x \pm
y} \Delta \!  ^{2} \!  \hat X \!  ^{-}_{x \pm y})^{- \! 
\frac{1}{2}}}{2} -1
\label{nmin}
\end{equation}    
and is a measure of the degree of entanglement.  Any photons in excess
of $\bar n_{{\rm min}}$ are not necessary for the entanglement and
contribute to the mixedness of the state.  The number of excess
photons $\bar n_{{\rm excess}}$ is just the total number of photons of
both beams minus the number required to sustain the entanglement, and 
can be written in terms of the average of the amplitude and phase 
quadrature variances of the two entangled beams
\begin{eqnarray}
\bar n_{{\rm excess}}  &=&  \bar n_{x} + \bar n_{y} - \bar n_{{\rm 
min}} \label{nexcess}\\
&=& (\Delta \!  ^{2} \!  \hat X^{+}_{x} \!  + \!  \Delta \!  ^{2} \! 
\hat X^{-}_{x} \!  + \!  \Delta \!  ^{2} \!  \hat X^{+}_{y} \!  + \! 
\Delta \!  ^{2} \!  \hat X^{-}_{y})/4 \!  - \!  \bar n_{{\rm min}} \! 
- \!  1 \nonumber
\end{eqnarray}    

\begin{figure}[b]
    \begin{center}
    \includegraphics[width=8.5cm]{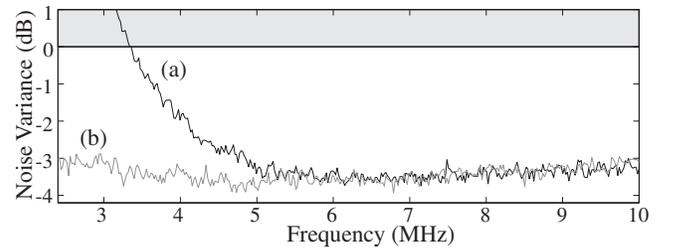}
    \end{center}
\vspace{-5mm} \caption{ Frequency spectra of (a) $\Delta \!  ^{2} \! 
\hat X \!^{+}_{x \pm y}$ and (b) $\Delta \!  ^{2} \!  \hat X \! 
^{-}_{x \pm y}$ for our entangled beams with no additional loss.}
    \label{DuanAP}
\end{figure}

\begin{figure}[t]
   \begin{center}
   \includegraphics[width=8.7cm]{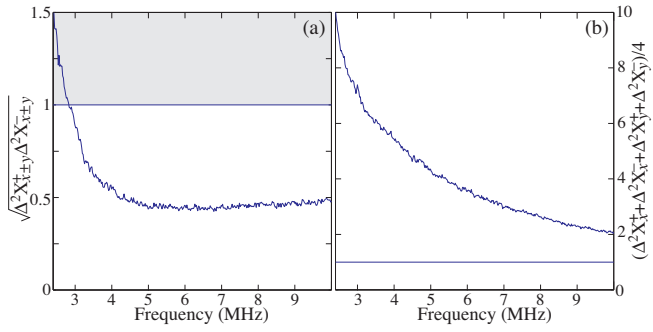}
   \end{center}
\vspace{-5mm} \caption{Spectra of (a) $\sqrt{\Delta^{2}\!  X_{x \pm
y}^{+} \Delta^{2}\!  X_{x \pm y}^{-}}$, and (b) the average of the
quadrature noise variances of the individual entangled
beams; with no additional loss.}
   \label{DuanandFluff}
\end{figure}

We characterized $\bar n_{{\rm min}}$ and $\bar n_{{\rm excess}}$ for
our entanglement over the frequency range from 2.5~MHz to 10~MHz. 
Measured spectra for $\Delta \!  ^{2} \!  \hat X \!^{+}_{x \pm y}$ and
$\Delta \!  ^{2} \!  \hat X \!  ^{-}_{x \pm y}$ are shown in
fig.~\ref{DuanAP}.  The amplitude quadrature measurement is
significantly degraded at low frequencies due to the resonant
relaxation oscillation of our laser.  Effectively this means that at
low frequencies we have a mixed state with many more photons present
than necessary for our entanglement.  At high frequencies both spectra
degrade due to the bandwidth limitations of our OPAs.
Fig.~\ref{DuanandFluff}(a) shows the resulting spectra for
$\sqrt{\Delta \!  ^{2} \!  \hat X \!^{+}_{x \pm y} \Delta \!  ^{2} \! 
\hat X \!  ^{-}_{x \pm y}}$, this spectra directly results in a
spectra for $\bar n_{{\rm min}}$.  We measured the amplitude and phase
quadrature variances of each entangled beam independently.  The
average of these measurements, which provides the total photon number
$\bar n_{x}\!+\!\bar n_{y}$, is shown in fig.~\ref{DuanandFluff}(b).

\begin{figure}[t]
   \begin{center}
   \includegraphics[width=8.7cm]{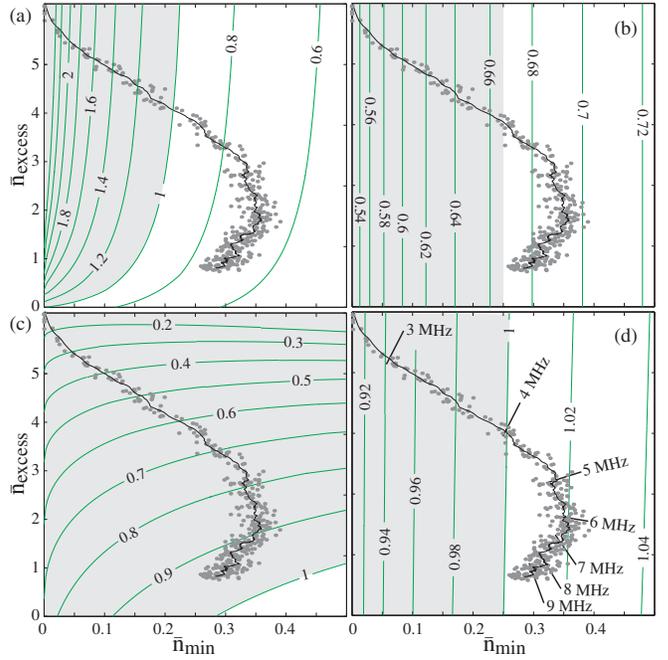}
   \end{center}
   \vspace{-2mm} 
   \vspace{-3mm} 
   \caption{Experimental results on photon number plots. 
   The contours on the plots are (a) EPR criterion, (b) Fidelity
   measure of teleportation, (c) and (d) ratio of dense-coding channel
   capacity to optimum squeezed channel capacity for mean photon
   numbers of 6.75 and 250, respectively.}
   \label{nQnexcess}
\end{figure}

$\bar n_{{\rm min}}$ and $\bar n_{{\rm excess}}$ were calculated from
eqs.~(\ref{nmin})~and~(\ref{nexcess}), and are plotted against each
other in fig.~\ref{nQnexcess}.  At low frequencies the number of
excess photons is large due to contributions from our lasers resonant
relaxation oscillation, and the number of photons required to maintain
the entanglement is small due to poor squeezing.  As the frequency
increases away from the relaxation oscillation the number of excess
photons decreases and the entanglement, and therefore $\bar n_{{\rm
min}}$ , improves.  At high frequencies the bandwidth of our OPAs
limit the amount of squeezing and consequently the number of photons
required to maintain the entanglement decreases.  The contours on each
of the four plots in fig.~\ref{nQnexcess} show the efficacy of
different protocols that the entanglement might be useful for.  The
details of how these contours were calculated will be included in a
future publication \cite{Bowen02B}.  Fig.~\ref{nQnexcess}(a) shows
efficacy contours for demonstration of the EPR-paradox as defined in
criterion~(\ref{VcvVcv}), the curving of the contours shows the
dependence of the criterion on mixedness.  This plot implies that for
our entanglement the optimum sideband frequency for observation of the
EPR-paradox is near 6.6~MHz.  Fig.~\ref{nQnexcess}(b) shows efficacy
contours for unity gain quantum teleportation \cite{CVtele}.  Here the
contours run vertically, with no dependence on the mixedness of the
entangled state, this result implies that optimum teleportation
results would be observed near 6.2~MHz.  Fig.~\ref{nQnexcess}(c)
shows contours of the channel capacity for densecoding when an average
of only 6.75 photons per bandwidth per time are allow in the
sidebands.  These contours show an extremely strong dependence on the
number of excess photons carried by the entanglement, this is because
every excess photon is one less that can be used to encode
information.  Accordingly, the optimum channel capacity would occur at
10~MHz where our entanglement is most pure.  Increasing the total
number of photons allowed in the sidebands can negate this problem
entirely \cite{Ralph02}.  In fig.~\ref{nQnexcess}(d) we show efficacy
contours for densecoding where 250 photons per bandwidth per time are
allowed in the sidebands.  In this case the contours are almost
vertical and show very little dependence on the number of excess
photons, resulting in an optimum channel capacity near 6.3~MHz.

In conclusion, we have produced strong, stably locked quadrature
entanglement.  It was characterized in terms of EPR and inseparability
criteria with optimum results of $\Delta^{2}\!  X_{x|y}^{+}
\Delta^{2}\!  X_{x|y}^{-} \!  = \!  0.58 \!  \pm \!  0.02$ and
$\sqrt{\Delta^{2}\!  X_{x \pm y}^{+} \Delta^{2}\!  X_{x \pm y}^{-}}\! 
= \!  0.44 \!  \pm \!  0.01$, respectively.  We observed qualitatively
and quantitatively different behaviour for the two criteria on the
introduction of controlled decoherence.  A more complete
characterization of entanglement can be achieved by considering both
the inseparability and the mixedness of the entanglement.  We
introduce a photon number plot that characterizes the entanglement in
terms of the number of photons necessary to maintain the entanglement,
which can be directly related to the inseparability criterion, and the
number of excess photons that contribute to mixedness.  We represent
our experimental results on such a diagram, and present efficacy
contour for the EPR criterion, and teleportation and densecoding
protocols.


\begin{thebibliography}{12}
%
\bibitem{Reid88}M.~D.~Reid and P.~D.~Drummond, \prl {\bf 60}, 2731
(1988); M.~D.~Reid, quant-ph/0112038.
%
\bibitem{Duan00} L-M.~Duan {\it et al.}, \prl {\bf 84}, 2722 (2000).
%
\bibitem{Bennett93} C.~H.~Bennett {\it et al.}, \prl {\bf 70}, 1895
(1993).
%
\bibitem{Bennett92} C.~H.~Bennett and S.~J.~Wiesner, \prl {\bf 69},
2881 (1992); S.~L.~Braunstein and H.~J.~Kimble, \pra {\bf 61},
042302 (2000).
%
\bibitem{Ralph02} T.~C.~Ralph and E.~H.~Huntington, quant-ph/0208117.
%
\bibitem{DiVincenzo95} D.~P.~DiVincenzo, Science {\bf 270}, 255
(1995).
%
\bibitem{Einstein35} A.~Einstein {\it et al.}, Phys.~Rev.  {\bf 47},
777 (1935).
%
\bibitem{Bowen02}W.~P.~Bowen {\it et al.}, \prl {\bf 88}, 093601
(2002); R.~Schnabel {\it et al.}, quant-ph/0208103.
%
\bibitem{conditional} See for example Z.~Y.~Ou {\it et al.}, \prl {\bf
68}, 3663 (1992); Yun~Zhang {\it et al.}, \pra {\bf 62}, 023813
(2000); and Ch.~Silberhorn {\it et al.}, \prl {\bf 86}, 4267 (2001).
%
\bibitem{CVtele}S.~L.~Braunstein and H.~J.~Kimble, \prl {\bf 80}, 869
(1998); and T.~C.~Ralph and P.~K.~Lam, \prl {\bf 81}, 5668 (1998).
%
\bibitem{Bowen02B}W.~P.~Bowen {\it et al.}, {\it Generation and
characterization of quadrature entanglement} yet to be written.
\end{thebibliography}
\end{document}